\documentclass[12pt,a4paper]{article}
\usepackage{latexsym,amsfonts,amssymb,epsfig,fleqn,euscript}
\usepackage{url}
\usepackage{color}
\definecolor{red,green,blue}{rgb}{0.5,0.5,0.7}
\definecolor{cyan,magenta,yellow,black}{cmyk}{0.5,0.5,0.5,1}

\def\R{{\rm I\mkern-2.4muR}} 

\def\C{\mathchoice{\rm C\kern-5.5pt\vrule width.07em height1.54ex depth0ex%
                   \kern4.7pt}
                  {\rm C\kern-5.5pt\vrule width.07em height1.52ex depth-.05ex%
                   \kern4.7pt}
                  {\scriptstyle\rm C\kern-4pt%
                   \vrule width 0.6pt height 4.7pt depth -0.8pt\kern3.4pt}%
                  {\scriptscriptstyle\rm C\kern-3.5pt%
                   \vrule width 0.4pt height 3.3pt depth -0.6pt\kern3.1pt}}
\def\S0S{\rho^{\R}}
\def\r0{\rho_0}

\def\S{{\mathfrak S}}

\def\Sum1{\mbox{\rm Sum1}}

\def\Dot1{\mbox{\rm Dot1}}

\def\R2n{{\bf {R}}^{2n}}

\def\R{{\bf {R}}}

\title{A classical interpretation of quantum mechanics and the measurement problem}
\author{Christian ~Jansson}

\begin{document}

\maketitle

\vspace*{0.5cm} Institute for Reliable Computing, Technical
University Hamburg--Harburg, Schwarzenbergstra{\ss}e 95, 21071
Hamburg, Germany, e-mail: jansson@tu-harburg.de, Fax: ++49 40
428782489.

\begin{center}
\vspace*{0.5cm} \noindent{\bf{Work in progress}}
\end{center}

\begin{abstract} In this paper a didactic  approach is described which immediately leads to an understanding of those postulates of quantum mechanics used most frequently in quantum computation. Moreover, an interpretation of quantum mechanics is presented which is motivated by retaining the point of view of classical mechanics as much as possible, and which is consistent with relativity theory. Everything can be written down in terms of well-known mathematical formulations that can be found in every textbook about quantum mechanics. Therefore, in this version, almost no formulas are used.

\end{abstract}

\section{Introduction}

During the last years I gave short courses in quantum computation to students studying electrical engineering and computer science. Most of these students have only a rudimentary knowledge in physics, but they have an intuitively understanding of classical mechanics. The postulates in quantum mechanics, the mathematical formalism, and the interpretation appears very confusing to them (and also to me). Therefore, I tried to write down classical mechanics in terms of the postulates of quantum mechanics from the point of view of duality (mainly I work in numerical analysis and optimization). Moreover, I tried to give all quantum mechanical quantities some realism.

Since I always had a bad feeling to introduce the collapse postulate only with the interpretation ``Shut up and Calculate'', in my last course I decided to use the
many world interpretation (being an interpretation of superposition) which introduces realism, and entanglement to be described in terms of different worlds. To me this seems to be a good didactic concept for understanding what happens in quantum mechanics. H.D. Zeh \footnote{Wozu braucht man Viele Welten in der Quantentheorie - auch f{\"u}r Nichtphysiker gedacht} did write a very impressive paper  about decoherence, many world interpretations, and realism. I used it in my last course on quantum computation, and it inspired me to write this manuscript which expresses my own conception as non-physicist.

This paper is intended in an inquiring manner. I am deeply grateful for every answer.

\section{Classical Derivation of the Postulates}

In duality theory a problem has two faces and two types of variables, the primal and the dual ones. Let us take as primal variables the canonical coordinates $q$  and the conjugate momenta $p$ of a particle; that is the primal state $(q,p)$ is in the linear phase space $M = {\bf {R}}^6$ of classical mechanics. Then one can identify these points with Dirac delta functions on $M \times M$, or equivalently as density matrices $\rho$ with index set $M$, which have on the diagonal exactly in position $((q,p),(q,p))$ a nonzero entry, say one, and are zero otherwise. These matrices are basis elements of the dual linear space of complex matrices, they are positive with trace one, and hence they are special density matrices corresponding to the idealized classical states (more precisely we should use classical phase space distributions like in Liouville's equation). These matrices describe the dual variables, and we obtain:

\begin{quote}

\emph{Postulate 1: (States)} A classical state is completely  described by a basis density matrix.

\end{quote}

A classical observable is described by a function $a(q,p)$ that is expressed in the dual space as a diagonal matrix $A$ with the function $a$ on the diagonal, yielding

\begin{quote}

\emph{Postulate 2: (Observable)} A classical observable is described by a a diagonal matrix $A$, and in the classical state $\rho$ we observe the sharp value $\langle A \rangle = trace(\rho A).$ Measurement does not change the state.

\end{quote}

The dynamics of the primal variables can be described, for example, by Hamilton's equations, and using Poisson brackets we obtain
\[
da / dt = \{ a,H \} + \partial a / \partial t,
\]
where $H$ is the Hamiltonian, $da / dt$ describes the dynamics of the observable, and $\partial a / \partial t$ is the classical derivative of $a(q(t),p(t),t)$.
This equation does not change, if it is written in terms of diagonal matrices in the dual space.

\begin{quote}

\emph{Postulate 3: (Dynamics)} The evaluation of a classical  observable $A$ is defined by Hamilton's equations using Poisson brackets.  The transition from one classical state to another one is described in the dual space by a permutation matrix $P$, yielding the special unitary transformation
$\rho' = P \rho P^T$.

\end{quote}

Rather straightforward is

\begin{quote}

\emph{Postulate 4: (Composition)} Two classical states, which correspond to a point in $M^2$, are represented as the tensor product of the corresponding basis density matrices in the dual space, yielding a basis density matrix with index set $M^2$.

\end{quote}

It follows that the postulates of quantum mechanics describe classical mechanics in a dual formulation by additionally selecting special density matrices, diagonal matrices as special hermitian matrices, and permutation matrices as special unitary matrices. Vice versa, allowing general density matrices, hermitian matrices, and replacing Hamilton's principle by the Heisenberg picture
\[
dA / dt = (i\hbar)^{-1} [ A,H ] + (\partial A / \partial t)_{classical}
\]
immediately yields quantum mechanics. Alternatively, also the Liouville equation and its analog in quantum mechanics, or the Schr{\"o}dinger picture may be used. Write $(i\hbar)^{-1}$ into the definition of the brackets yields exactly the same formula, but with different interpretation of the quantities. Observe that duality is only meaning in context with the notion and should not be confused with complementarity.

Summarizing, classical mechanics and quantum mechanics appear in the same outer shape.
Of course, using this entrance into the postulates does not give a deeper understanding of quantum mechanics for students. But, within a short period of time, they can work very naturally with the quantum postulates, and probably they do not despair at the beginning and later, as H.D. Zeh describes in his paper:

\begin{quote}
Tats{\"a}chlich ergab sich diese explizite Absage an eine "reale Quantenwelt"
aber aus ganz konkreten Konsistenzfragen, etwa ob das Elektron nun wirklich ein Teilchen
oder eine Welle sei. In der operationalistischen Realit{\"a}t der physikalischen Anwendung hat
sich diese pragmatische Haltung ausserordentlich bew{\"a}hrt, da sie sich nicht nur als hierf{\"u}r
ausreichend erwiesen hat, sondern die Beteiligten auch nicht von der Fortsetzung ihrer Arbeit
abhielt. (Ich kenne eine ganze Reihe ausgezeichneter junger Physiker, die durch die unl{\"o}sbar
erscheinenden Konsistenzprobleme der Quantentheorie v{\"o}llig blockiert wurden und dadurch
in ihrer beruflichen Karriere scheiterten.) Andererseits f{\"u}hrt sie aber zu einer negativen nat{\"u}rlichen
Auslese bez{\"u}glich m{\"o}glicher Fortschritte im Verst{\"a}ndnis der Theorie und damit unserer
Welt. Die Kopenhagener Deutung wird heute in den meisten Lehrb{\"u}chern als "Standardinterpretation"
(wenn es denn {\"u}berhaupt eine ist) bezeichnet, aber ohne dass ihre volle Bedeutung
und Ungeheuerlichkeit dabei hinreichend klargestellt wird. Das f{\"u}hrt dann zu den naiven
Fehlinterpretationen.
\end{quote}

However, looking once more into other interpretations (Kopenhagen, von Neumann, ensembles, decoherence, Feynman's path integrals, Bohm's interpretation, Quantum Logic, Algebras, etc.), I became very confused. Finally, the paradoxes appeared to me as a problem of time, the physical one but also my personal time. Nevertheless, I developed a version which in my opinion has much realism, can be understood in terms of classical mechanics, also from the point of view of engineers. It avoids many worlds and many minds, by introducing only two pictures (two worlds) of the universe. In some sense this interpretation comprises several faces of some of the well-known interpretations. It gives me a possible interpretation what may happen in quantum gravity.

\section{The GMc Interpretation} I call  this interpretation the ``Gravitation-Motion of Matter-Light with maximal speed c'' interpretation. It is based on the view of the two complementary particles photons and gravitons.

Photons are particles with zero rest mass and traveling (in vacuum) at the maximum speed of light c in each direction. They cannot be fixed, they have no position (i.e., the position is not defined), and their point of view is that everything is at rest. They are falling into the eyes of observers and generate pictures of the past which show the relativity of motion and actions as in a film. In other words, they produce a real film of the universe. The corresponding film reel consists of small proceeding pieces, one piece after the other described by an ordering, say $T$. Photons are used to define the space-time in relativity theory. For example, the Lorentz transformation expresses time in terms of speed of light, that is, in terms of particles. One advantage is the fact that Maxwell's equations (in vacuum) take the same form in any inertial coordinate system. In this perspective nothing is absolute; time, distance, speed, mass, e.t.c. have only a relative meaning, and every value can be measured depending on the state. Hence these quantities are observables. This film provided by the photons, I say shortly the p-picture, expresses our understanding of the world, and it is Einstein's point of view.

Gravitons are hypothetical elementary particles with zero motion mass. They cannot be put into motion (velocity is not defined), and they are fixed in points, say $X$. They are responsible for gravitational forces with directions, say in $V$. Since motion and velocity is not defined for gravitons, it follows that quantities like time and distance, well defined in the p-picture, have no meaning in their picture, which I call the g-picture.  They produce a film of prognosis for the universe, depending on the p-picture, and their film reel has also the ordering $T$. Visualized, they recognize the motion of matter in terms of the historical photos delivered from the p-picture. Their sense is that matter behaves like water, always moving and fluctuating, and when they are asked for a dynamics they can give at most probabilities for the future photos in the film; in other words they provide a film of probability densities of the universe. This picture is also Zenon's point of view. It yields, for example, to a precise description of the phrase that a particle is at the same time in two positions: In their picture, the particle can be in two positions.

Both films are also components of our consciousness, and we process our surroundings in form of an interplay of both films.
In my opinion, these different perspectives of both types are also responsible for the paradoxes in relativity theory and quantum mechanics. They appear in the same way in both theories.

\section{The Postulates}
In this section I briefly describe the frame of the GMc-interpretation. The ingredients are:

(1) There is a three-dimensional space of points, say $X$.

(2) There is a three-dimensional space of vectors representing directions, say $V$.

(3) There is an absolute time $T$ defining an ordering of the universe (which should not be confused with our measured time in relativity theory).

(4) The gravitational part of the universe is described by the family $(X,V,T)$ with states $(x,v,T)$, and the ordering of states $(x_1,v_1,T_1) \ge (x_2,v_2,T_2)$ if $T_1 \ge T_2$.

A state describes the degree of freedom of a particle, or in other words, the independent variables position $x$ and velocity $v$. Additionally, $T$ describes an ordering where you cannot go back. If there are no external forces, then the particle moves w.r.t. $T$ in $X$ on a line in direction $v$, and $v$ does not change.

The motion of matter: Photons are always in states $(\infty,c,T)$, and gravitons are always in states $(x,\infty,T)$, where $\infty$ means undefined.
All other particles live in this six-dimensional universe with well-defined states $(x,v,T)$ denoting position and velocity, and describing the independent variables.
The time $t$ is an observable (a variable depending on two states $((x,v,T),(x',v',T))$), and it means that two particles are in two positions $x,x'$ with two speeds $v,v'$. This time is defined according to the formalism in relativity theory, and it is expressed, for example, in the Lorentz transformation. The other observables, mass, position, velocity, momentum, kinetic energy w.r.t. $V$, potential energy w.r.t $X$, etc., can now be expressed in the usual way in terms of the relative time $t$. For photons and gravitons the relative time is undefined, since they have no defined position or speed. However, their states are ordered w.r.t. to $T$.

Einstein's field equation describe this universe w.r.t. the observable time and the p-picture (everything is at rest), yielding the well-know curved space-time. This equation can be expressed also in terms of  $X$ and $V$, since each state $(x(t),v(t),t)$ of the curved space-time yields a state $(x,v,T)$ w.r.t. the ordering $T$. Then the field equation should look similar to Maxwell's equations: There is the field of motion and the gravitational field, and the field of motion is determined by the mass-energy distribution, etc. Photons should move in the vacuum on straight lines in this framework.
For describing experiments in a laboratory, classical mechanics in form of corresponding differential equations or principles of action, possibly enriched with laws of thermodynamics, may be sufficient.
The set of all interacting particles (containing all observers and all measurement devices) can be viewed as a sea of water being always in motion and fluctuation from the point of view of the gravitons, and being always at rest for the photons.

Schr{\"o}dingers equation describes this universe in the g-picture (everything is in motion expressed by the historical photos, and one can only make prognosis for the future), yielding the complex probability amplitude or alternatively the probability density of the universe.

Using this classical setting of the world, the major postulates of this interpretation read as follows:

\begin{quote}

\emph{Postulate 1: (Classical Principle)} Particles can be only in classical states $(x,v,T)$. These states are called the possible states.

\end{quote}

This means explicitly that particles are no waves or fields, there is no superposition principle for a particle, and we have only one universe. We write $(x,v,T)_m$ to denote a state of a certain particle with mass $m$.

\begin{quote}

\emph{Postulate 2: (Particle Motion)} All particles can be only in states $(x,v,T)_M$ determined by relativity theory (occasionally classical mechanics and thermodynamics). These states are called the feasible states.
\end{quote}

The set $p(X,V,T)_M$ describes all feasible states $(x,v,T)_m$ for all particles of the universe at ordering time $T$,
and is called a motion-photo. The ordered set of all motion-photos is called the motion-film.

\begin{quote}

\emph{Postulate 3 : (Quantum Prognosis)} The probability density of the universe $\rho$ is defined on $p(X,V,T)_M$ at ordering time T and satisfies the Schr{\"o}dinger equation; that is, the passage from $T_0$ to $T_1$ is defined by a unitary transformation.
\end{quote}

The probability density may be also defined on all possible states, but then $\rho$ is nonzero only for feasible states.
The function $\rho(p(X,V,T)_M)$ describes the probabilities density for all feasible states of all particles at ordering time $T$, and is called a probability-photo. The ordered set of all probabilities photos is called the probability-film. Both types of photos and films are complementary.

\begin{quote}

\emph{Postulate 4 : (Dynamics)} The dynamics is a rotating process, ordered by $T$, between both complementary photos of the films with the smallest amount of action $\hbar$.
\end{quote}

Hence, motion is described in terms of action of the particles according to the particle motion, yielding new feasible states. This implies a change of the probability density $\rho$. This affects motion, and this affects $\rho$. Hence, the dynamics is an interchanging reaction between both pictures and may be described by an operator which behaves unitary w.r.t. $\rho((X,V,T)_M)$ and nonlinear w.r.t. $(X,V,T)_M$.

Measurements are always done in the motion-film (p-picture), these are the photos where everything is at rest. In this sense measurement does not change the state. Probabilities are always expressed in the probability film (g-picture). Only probabilities are transformed unitary, not states. Particles are never in superposition. Hence, in this interpretation probabilities remain reversible, but due to the interplay in the dynamics irreversibility is introduced. It follows that the particle-wave dualism is dissolved by a nonlinear motion-film and a unitary probability-film, with the advantage that particles remain particles, measurement is done in each moment by the motion photos, and thus does not change states.

In this setting, building up a measurement already apparatus provides the probability density and interference patterns, {\it before a particle is in the experiment}. The apparatus is fluctuating w.r.t. the gravitons. Superpositions of different states of a particle are not real, and they are not required in this interpretation, every uncertainty is described in the probability density $\rho$. The complex wave function is only a mathematical quantity: Required to compute the probability density, and for the purpose of convenience, for example to work with vectors instead of matrices.
At the time when the particle is in the device, for example in the double slit, a path (slit) is chosen which is reflected in $\rho$. This is decoherence in the GMc interpretation. It should not be confused with the well-known decoherence, which assumes that the particle has wave properties and is in superposition.

If there is no device where the particle can choose different paths or is scattered, we have  also the interplay of the pictures, but the particle proceeds classical, since the relevant part of the probability density now has sharp values. In particular, then this model passes over to classical mechanics and relativity theory.

The interpretation of the position-momentum uncertainty in this model is as follows: In the p-picture we obtain photos where the universe is at rest. Hence, the position of a particle is known exactly, but nothing can be said about the momentum. A sequence of these photos, with the interchange of the pictures, necessarily yields an uncertainty of position and momentum at fixed $T$. The interpretation of the time-energy uncertainty reads as follows: In the g-picture we obtain probability photos about position and momentum. The time and energy is not known, but can be estimated from a sequence of these probability-photos. Important is that this is an estimation which cannot be improved arbitrarily, that is the estimation is bounded w.r.t. $\hbar$.

This interpretation also applies to larger objects in the manner that a camel cannot go through the eye of a needle. In other words, if the fluctuations of the measurement apparatus w.r.t. gravitons are small compared to the size of the object (particle), interference does not occur. This should imply that the Heisenberg cut can be made more explicitly.

In this dynamics, the connection between gravitation and quantum mechanics is given by the feasible states, and I don't see a contradiction. Are there any? In my opinion both theories are complementary, that is, all rules from both theories are necessary. It explains why in quantum mechanics particles always arrive as particles in feasible states, and a collapse postulate is not required. We have only one necessarily coupled dynamics: Postulate 2 describes the feasible states in a relative manner: at fixed T we obtain a  p-photo, and Postulate 3 gives prediction for the future. This is my understanding of quantum gravity.

\section{Experiments}
In this section some experiments are described, only very briefly, without using any mathematics, instead taking more the point of view of an engineer.

First, the key experiment, the double slit experiment: Feynman was fond of saying that all of quantum mechanics can be gleaned from carefully thinking through the implications of this single experiment. If only one slit is open, in the g-picture, the whole apparatus and environment is in fluctuation (evidenced by the p-photos) yielding an uncertainty in momentum, if a particle passes the slit. This uncertainty is expressed in $\rho$, and  it is already known before a particle is in the experiment. Hence, in the g-picture, the measurement apparatus  is responsible for the uncertainty and probabilities, and not the particle. If the slit becomes smaller the position is more precisely, and the fluctuations of the apparatus in the g-picture become larger relative to the diameter of the slit,  yielding larger disturbances of the momentum. This changes the uncertainty and the probability density, accordingly, before a particle is in the slit.
In this interpretation, there is no difference between ensembles and single particles.

If both slits are open, then, in the g-picture, the fluctuations (evidenced by the p-photos) generates two vibrating slits yielding the well-known interference pattern with the corresponding probability density, before the particle is in the experiment. The particle is always a particle, and it is not a wave or split in some sense,
like in many worlds. If the particle is entering one slit, then decoherence has appeared and is reflected in $\rho$ in the usual way. It is not clear at that moment which slit is chosen by the particle. However, immediately hereon, the particle is now on one of the paths leading to the detectors on the wall behind.  The p-photos (if its dark then the surrounding does the job of the photons) gives information on which path the particle is. This is reflected in the probability density, and the measurement at the wall is only the confirmation of things which already are known and which happened in the past.

Finally, let us now discuss the case where both slits are open, and we measure through which slit the particle passes, for example by using photons and detectors. Then the moving photons generate a multiple slit where the particle will be scattered. Hence, we obtain an interference pattern close to the two slits, which is reflected in $\rho$ before the particle is in the apparatus. Then, if the particle is in the experiment, there happens the same as above. But since the interference pattern is close to the slits it is known through which slit the particle has passed, and this is almost equivalent (w.r.t. the probability amplitude) to the case where the other slit is closed.

Everything is described completely in terms of classical logic and classical mechanics. In the GMc interpretation both pictures exist at each absolute time $T$ together with their interplay. The usual interpretation in quantum mechanics (for example in the double slit experiment) is that the measurement devices and the observer are fixed in position, and hence the particle must be a wave or field. Here, the observer, the measurement device and the whole universe, with exception of the gravitons,  move and fluctuate.

Next, let us look at the Mach-Zehnder-interferometer. There, we have a lower and an upper trajectory. The whole apparatus with environment fluctuates (say like a wave) in the g-picture. The symmetry of this apparatus yields: The lower part of this fluctuating wave passes the first half-silvered mirror, is reflected at the mirror (this is not relevant), and then is reflected at the second half-silvered mirror. The upper part of the fluctuating wave has the reverse destiny. The fluctuation is in both directions. Hence, if there is coming something in from the left, it has also to go out to the left.
Building up the interferometer generates the correct $\rho$, which is known before a particle enters the interferometer. Now, if we put a bomb in one of the trajectories, the fluctuating wave is interrupted yielding the well-known observation of Elitzur and Vaidman. This is an explanation of interactive-free measurements: everything is already measured before a particle is in the apparatus.

I want to describe briefly entanglement in the case of the thought experiment  ``Schr{\"o}dinger's cat''. This is very simple in the GMc framework. The cat lives as long as there is no decay. The probability density in the g-picture contains the classical probability w.r.t. the decay. There is no superposition of dead and alive. If the decay happens, then, due to the interplay, the cat will be dead, in the p-picture the cat lies dead on the bottom, and in the g-picture the probability that the cat is dead is exactly one. This explains also Wigner's friend.

The principle of locality is that distant objects cannot have direct influence on one another. This principle is fulfilled in the GMc interpretation. Due to the interplay of both pictures, immediately at the very beginning of the experiment, like above, it is clear which classical state is chosen. The decision which classical state of an entangled Bell state is chosen happens at the very beginning, just like in the many world interpretation, but without splitting the world. The interplay of both pictures replaces the split in many worlds. Hence, we have locality like in the Kopenhagener interpretation or consistent histories. But we also have realism.
This does not contradict Bell's inequality since we have two different pictures.

One remark to the twin paradox. In the GMc interpretation the situation would be as follows: The clock of the fast moving twin  measures another time
than the clock on the earth. But w.r.t. the ordering $T$, the photos and their interplay appear for both in the same way. Hence, they should always have the same biological age. Time in relativity theory is only an observable.

Coming back to quantum computation: The point of view that one has to build an interference apparatus, and then, afterwards, one particle is put into the computer, is completely contrary to the view where the particles are superposed. This mental image may help for understanding quantum computation and for building quantum computers.

\section{Conclusion}
I am aware of the fact that this interpretation yields a non-familiar conception of motion and time. The already developed mathematical formulations can be used, however, it avoids many worlds and the determinism that all parts of a decision are realized, it is in the sense of Ockham, it can be understand in terms of classical mechanics, and, what is most important to me, provides freedom: The smallest degree of freedom is contained in Planck's constant. I think that quantum electrodynamics can be handled in a similar way.

\bibliography{H:/bib/extern,H:/bib/ti3}

\end{document}